%
%
%
\magnification=1200      
\overfullrule=0pt        
\hsize=6truein           
\vsize=8.5truein         
%
%
\catcode`\@=11
\newif\iffirstpage@\firstpage@true
\def\plainoutput{\shipout\vbox{%
 \iffirstpage@\global\firstpage@false
  \pagebody\makefootline%
 \else\makeheadline\pagebody
 \fi}%
 \advancepageno\ifnum\outputpenalty>-\@MM\else\dosupereject\fi}
\catcode`@=12
%
%
\count253=0
\count254=0
\def\bibitem{\advance\count253 by 1\item{[\number\count253]}}
%
%
\font\teneuf=eufm10
\font\seveneuf=eufm7
\font\fiveeuf=eufm5
\newfam\euffam
\textfont\euffam=\teneuf
\scriptfont\euffam=\seveneuf
\scriptscriptfont\euffam=\fiveeuf

%
%
\font\tenbi=cmmib10
\font\sevenbi=cmmib7
\font\fivebi=cmmib5
\newfam\bifam
\textfont\bifam=\tenbi
\scriptfont\bifam=\sevenbi
\scriptscriptfont\bifam=\fivebi
\def\bi#1{{\fam\bifam#1}}
%
%
\def\balpha{{\mathchar"090B}}
\def\bbeta{{\mathchar"090C}}

\def\btheta{{\mathchar"0923}}
\def\biota{{\mathchar"0913}}

\def\bxi{{\mathchar"0918}}

\def\bphi{{\mathchar"091E}}
\def\bvarphi{{\mathchar"0927}}
\def\bpsi{{\mathchar"0920}}
\def\bomega{{\mathchar"0921}}
%
%

\font\tenbbb=msbm10
\font\sevenbbb=msbm7
\font\fivebbb=msbm5
\newfam\bbbfam
\textfont\bbbfam=\tenbbb
\scriptfont\bbbfam=\sevenbbb
\scriptscriptfont\bbbfam=\fivebbb
\def\Bbb#1{{\fam\bbbfam#1}}
%
%
\font\tit=cmbx12 scaled\magstep 1
\font\abst=cmti8
\font\abstb=cmbx8

\font\sevenit=cmti7

\font\eightrm=cmr8
\font\eighti=cmmi8
\font\boldsymbols=cmbsy10
%
%
\def\bold#1{{\bf #1}}
%
%
\def\centre#1{\vbox{\rightskip=0pt plus1fill\leftskip=0pt plus1fill #1}}
\def\mynumsec#1{\advance\count254 by 1%
                \bigskip{\noindent\bf\number\count254. #1}%
                \par\nobreak\noindent\ignorespaces}
\def\mysec#1{\bigskip{\noindent\bf#1}\par\nobreak\noindent\ignorespaces}
\long\def\abstract#1{\bigskip\medskip
\vbox{\par\leftskip=30truept\rightskip=30truept\baselineskip=0.8%
      \baselineskip\noindent{\abstb Abstract. }\abst #1}\parindent=15truept}
%
%
\edef\thinlines{\the\catcode`@ }%
\catcode`@ = 11
\let\@oldatcatcode = \thinlines
\edef\@oldandcatcode{\the\catcode`& }%
\catcode`& = 11
\def\&whilenoop#1{}%
\def\&whiledim#1\do #2{\ifdim #1\relax#2\&iwhiledim{#1\relax#2}\fi}%
\def\&iwhiledim#1{\ifdim #1\let\&nextwhile=\&iwhiledim
        \else\let\&nextwhile=\&whilenoop\fi\&nextwhile{#1}}%
\newif\if&negarg
\newdimen\&wholewidth
\newdimen\&halfwidth
\font\tenln=line10
\def\thinlines{\let\&linefnt\tenln \let\&circlefnt\tencirc
  \&wholewidth\fontdimen8\tenln \&halfwidth .5\&wholewidth}%
\def\thicklines{\let\&linefnt\tenlnw \let\&circlefnt\tencircw
  \&wholewidth\fontdimen8\tenlnw \&halfwidth .5\&wholewidth}%
\def\drawline(#1,#2)#3{\&xarg #1\relax \&yarg #2\relax \&linelen=#3\relax
  \ifnum\&xarg =0 \&vline \else \ifnum\&yarg =0 \&hline \else \&sline\fi\fi}%
\def\&sline{\leavevmode
  \ifnum\&xarg< 0 \&negargtrue \&xarg -\&xarg \&yyarg -\&yarg
  \else \&negargfalse \&yyarg \&yarg \fi
  \ifnum \&yyarg >0 \&tempcnta\&yyarg \else \&tempcnta -\&yyarg \fi
  \ifnum\&tempcnta>6 \&badlinearg \&yyarg0 \fi
  \ifnum\&xarg>6 \&badlinearg \&xarg1 \fi
  \setbox\&linechar\hbox{\&linefnt\&getlinechar(\&xarg,\&yyarg)}%
  \ifnum \&yyarg >0 \let\&upordown\raise \&clnht\z@
  \else\let\&upordown\lower \&clnht \ht\&linechar\fi
  \&clnwd=\wd\&linechar
  \&whiledim \&clnwd <\&linelen \do {%
    \&upordown\&clnht\copy\&linechar
    \advance\&clnht \ht\&linechar
    \advance\&clnwd \wd\&linechar
  }%
  \advance\&clnht -\ht\&linechar
  \advance\&clnwd -\wd\&linechar
  \&tempdima\&linelen\advance\&tempdima -\&clnwd
  \&tempdimb\&tempdima\advance\&tempdimb -\wd\&linechar
  \hskip\&tempdimb \multiply\&tempdima \@m
  \&tempcnta \&tempdima \&tempdima \wd\&linechar \divide\&tempcnta \&tempdima
  \&tempdima \ht\&linechar \multiply\&tempdima \&tempcnta
  \divide\&tempdima \@m
  \advance\&clnht \&tempdima
  \ifdim \&linelen <\wd\&linechar \hskip \wd\&linechar
  \else\&upordown\&clnht\copy\&linechar\fi}%
\def\&hline{\vrule height \&halfwidth depth \&halfwidth width \&linelen}%
\def\&getlinechar(#1,#2){\&tempcnta#1\relax\multiply\&tempcnta 8
  \advance\&tempcnta -9 \ifnum #2>0 \advance\&tempcnta #2\relax\else
  \advance\&tempcnta -#2\relax\advance\&tempcnta 64 \fi
  \char\&tempcnta}%
\def\drawvector(#1,#2)#3{\&xarg #1\relax \&yarg #2\relax
  \&tempcnta \ifnum\&xarg<0 -\&xarg\else\&xarg\fi
  \ifnum\&tempcnta<5\relax \&linelen=#3\relax
    \ifnum\&xarg =0 \&vvector \else \ifnum\&yarg =0 \&hvector
    \else \&svector\fi\fi\else\&badlinearg\fi}%
\def\&hvector{\ifnum\&xarg<0 \rlap{\&linefnt\&getlarrow(1,0)}\fi \&hline
  \ifnum\&xarg>0 \llap{\&linefnt\&getrarrow(1,0)}\fi}%
\def\&vvector{\ifnum \&yarg <0 \&downvector \else \&upvector \fi}%
\def\&svector{\&sline
  \&tempcnta\&yarg \ifnum\&tempcnta <0 \&tempcnta=-\&tempcnta\fi
  \ifnum\&tempcnta <5
    \if&negarg\ifnum\&yarg>0
      \llap{\lower\ht\&linechar\hbox to\&linelen{\&linefnt
        \&getlarrow(\&xarg,\&yyarg)\hss}}\else
      \llap{\hbox to\&linelen{\&linefnt\&getlarrow(\&xarg,\&yyarg)\hss}}\fi
    \else\ifnum\&yarg>0
      \&tempdima\&linelen \multiply\&tempdima\&yarg
      \divide\&tempdima\&xarg \advance\&tempdima-\ht\&linechar
      \raise\&tempdima\llap{\&linefnt\&getrarrow(\&xarg,\&yyarg)}\else
      \&tempdima\&linelen \multiply\&tempdima-\&yarg
      \divide\&tempdima\&xarg
      \lower\&tempdima\llap{\&linefnt\&getrarrow(\&xarg,\&yyarg)}\fi\fi
  \else\&badlinearg\fi}%
\def\&getlarrow(#1,#2){\ifnum #2 =\z@ \&tempcnta='33\else
\&tempcnta=#1\relax\multiply\&tempcnta \sixt@@n \advance\&tempcnta
-9 \&tempcntb=#2\relax\multiply\&tempcntb \tw@
\ifnum \&tempcntb >0 \advance\&tempcnta \&tempcntb\relax
\else\advance\&tempcnta -\&tempcntb\advance\&tempcnta 64
\fi\fi\char\&tempcnta}%
\def\&getrarrow(#1,#2){\&tempcntb=#2\relax
\ifnum\&tempcntb < 0 \&tempcntb=-\&tempcntb\relax\fi
\ifcase \&tempcntb\relax \&tempcnta='55 \or
\ifnum #1<3 \&tempcnta=#1\relax\multiply\&tempcnta
24 \advance\&tempcnta -6 \else \ifnum #1=3 \&tempcnta=49
\else\&tempcnta=58 \fi\fi\or
\ifnum #1<3 \&tempcnta=#1\relax\multiply\&tempcnta
24 \advance\&tempcnta -3 \else \&tempcnta=51\fi\or
\&tempcnta=#1\relax\multiply\&tempcnta
\sixt@@n \advance\&tempcnta -\tw@ \else
\&tempcnta=#1\relax\multiply\&tempcnta
\sixt@@n \advance\&tempcnta 7 \fi\ifnum #2<0 \advance\&tempcnta 64 \fi
\char\&tempcnta}%
\def\&vline{\ifnum \&yarg <0 \&downline \else \&upline\fi}%
\def\&upline{\hbox to \z@{\hskip -\&halfwidth \vrule width \&wholewidth
   height \&linelen depth \z@\hss}}%
\def\&downline{\hbox to \z@{\hskip -\&halfwidth \vrule width \&wholewidth
   height \z@ depth \&linelen \hss}}%
\def\&upvector{\&upline\setbox\&tempboxa\hbox{\&linefnt\char'66}\raise
     \&linelen \hbox to\z@{\lower \ht\&tempboxa\box\&tempboxa\hss}}%
\def\&downvector{\&downline\lower \&linelen
      \hbox to \z@{\&linefnt\char'77\hss}}%
\def\&badlinearg{\errmessage{Bad \string\arrow\space argument.}}%
\thinlines
\countdef\&xarg     0
\countdef\&yarg     2
\countdef\&yyarg    4
\countdef\&tempcnta 6
\countdef\&tempcntb 8
\dimendef\&linelen  0
\dimendef\&clnwd    2
\dimendef\&clnht    4
\dimendef\&tempdima 6
\dimendef\&tempdimb 8
\chardef\@arrbox    0
\chardef\&linechar  2
\chardef\&tempboxa  2
\let\lft^%
\let\rt_%
\newif\if@pslope
\def\@findslope(#1,#2){\ifnum#1>0
  \ifnum#2>0 \@pslopetrue \else\@pslopefalse\fi \else
  \ifnum#2>0 \@pslopefalse \else\@pslopetrue\fi\fi}%
\def\generalsmap(#1,#2){\getm@rphposn(#1,#2)\plnmorph\futurelet\next\addm@rph}%
\def\sline(#1,#2){\setbox\@arrbox=\hbox{\drawline(#1,#2){\sarrowlength}}%
  \@findslope(#1,#2)\d@@blearrfalse\generalsmap(#1,#2)}%
\def\arrow(#1,#2){\setbox\@arrbox=\hbox{\drawvector(#1,#2){\sarrowlength}}%
  \@findslope(#1,#2)\d@@blearrfalse\generalsmap(#1,#2)}%
\newif\ifd@@blearr
\def\bisline(#1,#2){\@findslope(#1,#2)%
  \if@pslope \let\@upordown\raise \else \let\@upordown\lower\fi
  \getch@nnel(#1,#2)\setbox\@arrbox=\hbox{\@upordown\@vchannel
    \rlap{\drawline(#1,#2){\sarrowlength}}%
      \hskip\@hchannel\hbox{\drawline(#1,#2){\sarrowlength}}}%
  \d@@blearrtrue\generalsmap(#1,#2)}%
\def\biarrow(#1,#2){\@findslope(#1,#2)%
  \if@pslope \let\@upordown\raise \else \let\@upordown\lower\fi
  \getch@nnel(#1,#2)\setbox\@arrbox=\hbox{\@upordown\@vchannel
    \rlap{\drawvector(#1,#2){\sarrowlength}}%
      \hskip\@hchannel\hbox{\drawvector(#1,#2){\sarrowlength}}}%
  \d@@blearrtrue\generalsmap(#1,#2)}%
\def\adjarrow(#1,#2){\@findslope(#1,#2)%
  \if@pslope \let\@upordown\raise \else \let\@upordown\lower\fi
  \getch@nnel(#1,#2)\setbox\@arrbox=\hbox{\@upordown\@vchannel
    \rlap{\drawvector(#1,#2){\sarrowlength}}%
      \hskip\@hchannel\hbox{\drawvector(-#1,-#2){\sarrowlength}}}%
  \d@@blearrtrue\generalsmap(#1,#2)}%
\newif\ifrtm@rph
\def\@shiftmorph#1{\hbox{\setbox0=\hbox{$\scriptstyle#1$}%
  \setbox1=\hbox{\hskip\@hm@rphshift\raise\@vm@rphshift\copy0}%
  \wd1=\wd0 \ht1=\ht0 \dp1=\dp0 \box1}}%
\def\@hm@rphshift{\ifrtm@rph
  \ifdim\hmorphposnrt=\z@\hmorphposn\else\hmorphposnrt\fi \else
  \ifdim\hmorphposnlft=\z@\hmorphposn\else\hmorphposnlft\fi \fi}%
\def\@vm@rphshift{\ifrtm@rph
  \ifdim\vmorphposnrt=\z@\vmorphposn\else\vmorphposnrt\fi \else
  \ifdim\vmorphposnlft=\z@\vmorphposn\else\vmorphposnlft\fi \fi}%
\def\addm@rph{\ifx\next\lft\let\temp=\lftmorph\else
  \ifx\next\rt\let\temp=\rtmorph\else\let\temp\relax\fi\fi \temp}%
\def\plnmorph{\dimen1\wd\@arrbox \ifdim\dimen1<\z@ \dimen1-\dimen1\fi
  \vcenter{\box\@arrbox}}%
\def\lftmorph\lft#1{\rtm@rphfalse \setbox0=\@shiftmorph{#1}%
  \if@pslope \let\@upordown\raise \else \let\@upordown\lower\fi
  \llap{\@upordown\@vmorphdflt\hbox to\dimen1{\hss
    \llap{\box0}\hss}\hskip\@hmorphdflt}\futurelet\next\addm@rph}%
\def\rtmorph\rt#1{\rtm@rphtrue \setbox0=\@shiftmorph{#1}%
  \if@pslope \let\@upordown\lower \else \let\@upordown\raise\fi
  \llap{\@upordown\@vmorphdflt\hbox to\dimen1{\hss
    \rlap{\box0}\hss}\hskip-\@hmorphdflt}\futurelet\next\addm@rph}%
\def\getm@rphposn(#1,#2){\ifd@@blearr \dimen@\morphdist \advance\dimen@ by
  .5\channelwidth \@getshift(#1,#2){\@hmorphdflt}{\@vmorphdflt}{\dimen@}\else
  \@getshift(#1,#2){\@hmorphdflt}{\@vmorphdflt}{\morphdist}\fi}%
\def\getch@nnel(#1,#2){\ifdim\hchannel=\z@ \ifdim\vchannel=\z@
    \@getshift(#1,#2){\@hchannel}{\@vchannel}{\channelwidth}%
    \else \@hchannel\hchannel \@vchannel\vchannel \fi
  \else \@hchannel\hchannel \@vchannel\vchannel \fi}%
\def\@getshift(#1,#2)#3#4#5{\dimen@ #5\relax
  \&xarg #1\relax \&yarg #2\relax
  \ifnum\&xarg<0 \&xarg -\&xarg \fi
  \ifnum\&yarg<0 \&yarg -\&yarg \fi
  \ifnum\&xarg<\&yarg \&negargtrue \&yyarg\&xarg \&xarg\&yarg \&yarg\&yyarg\fi
  \ifcase\&xarg \or
    \ifcase\&yarg
      \dimen@i \z@ \dimen@ii \dimen@ \or
      \dimen@i .7071\dimen@ \dimen@ii .7071\dimen@ \fi \or
    \ifcase\&yarg
      \or
      \dimen@i .4472\dimen@ \dimen@ii .8944\dimen@ \fi \or
    \ifcase\&yarg
      \or
      \dimen@i .3162\dimen@ \dimen@ii .9486\dimen@ \or
      \dimen@i .5547\dimen@ \dimen@ii .8321\dimen@ \fi \or
    \ifcase\&yarg
      \or
      \dimen@i .2425\dimen@ \dimen@ii .9701\dimen@ \or\or
      \dimen@i .6\dimen@ \dimen@ii .8\dimen@ \fi \or
    \ifcase\&yarg
      \or
      \dimen@i .1961\dimen@ \dimen@ii .9801\dimen@ \or
      \dimen@i .3714\dimen@ \dimen@ii .9284\dimen@ \or
      \dimen@i .5144\dimen@ \dimen@ii .8575\dimen@ \or
      \dimen@i .6247\dimen@ \dimen@ii .7801\dimen@ \fi \or
    \ifcase\&yarg
      \or
      \dimen@i .1645\dimen@ \dimen@ii .9864\dimen@ \or\or\or\or
      \dimen@i .6402\dimen@ \dimen@ii .7682\dimen@ \fi \fi
  \if&negarg \&tempdima\dimen@i \dimen@i\dimen@ii \dimen@ii\&tempdima\fi
  #3\dimen@i\relax #4\dimen@ii\relax }%
\catcode`\&=4
\def\generalhmap{\futurelet\next\@generalhmap}%
\def\@generalhmap{\ifx\next^ \let\temp\generalhm@rph\else
  \ifx\next_ \let\temp\generalhm@rph\else \let\temp\m@kehmap\fi\fi \temp}%
\def\generalhm@rph#1#2{\ifx#1^
    \toks@=\expandafter{\the\toks@#1{\rtm@rphtrue\@shiftmorph{#2}}}\else
    \toks@=\expandafter{\the\toks@#1{\rtm@rphfalse\@shiftmorph{#2}}}\fi
  \generalhmap}%
\def\m@kehmap{\mathrel{\smash{\the\toks@}}}%
\def\mapright{\toks@={\mathop{\vcenter{\smash{\drawrightarrow}}}\limits}%
  \generalhmap}%
\def\mapleft{\toks@={\mathop{\vcenter{\smash{\drawleftarrow}}}\limits}%
  \generalhmap}%
\def\bimapright{\toks@={\mathop{\vcenter{\smash{\drawbirightarrow}}}\limits}%
  \generalhmap}%
\def\bimapleft{\toks@={\mathop{\vcenter{\smash{\drawbileftarrow}}}\limits}%
  \generalhmap}%
\def\adjmapright{\toks@={\mathop{\vcenter{\smash{\drawadjrightarrow}}}\limits}%
  \generalhmap}%
\def\adjmapleft{\toks@={\mathop{\vcenter{\smash{\drawadjleftarrow}}}\limits}%
  \generalhmap}%
\def\hline{\toks@={\mathop{\vcenter{\smash{\drawhline}}}\limits}%
  \generalhmap}%
\def\bihline{\toks@={\mathop{\vcenter{\smash{\drawbihline}}}\limits}%
  \generalhmap}%
\def\drawrightarrow{\hbox{\drawvector(1,0){\harrowlength}}}%
\def\drawleftarrow{\hbox{\drawvector(-1,0){\harrowlength}}}%
\def\drawbirightarrow{\hbox{\raise.5\channelwidth
  \hbox{\drawvector(1,0){\harrowlength}}\lower.5\channelwidth
  \llap{\drawvector(1,0){\harrowlength}}}}%
\def\drawbileftarrow{\hbox{\raise.5\channelwidth
  \hbox{\drawvector(-1,0){\harrowlength}}\lower.5\channelwidth
  \llap{\drawvector(-1,0){\harrowlength}}}}%
\def\drawadjrightarrow{\hbox{\raise.5\channelwidth
  \hbox{\drawvector(-1,0){\harrowlength}}\lower.5\channelwidth
  \llap{\drawvector(1,0){\harrowlength}}}}%
\def\drawadjleftarrow{\hbox{\raise.5\channelwidth
  \hbox{\drawvector(1,0){\harrowlength}}\lower.5\channelwidth
  \llap{\drawvector(-1,0){\harrowlength}}}}%
\def\drawhline{\hbox{\drawline(1,0){\harrowlength}}}%
\def\drawbihline{\hbox{\raise.5\channelwidth
  \hbox{\drawline(1,0){\harrowlength}}\lower.5\channelwidth
  \llap{\drawline(1,0){\harrowlength}}}}%
\def\generalvmap{\futurelet\next\@generalvmap}%
\def\@generalvmap{\ifx\next\lft \let\temp\generalvm@rph\else
  \ifx\next\rt \let\temp\generalvm@rph\else \let\temp\m@kevmap\fi\fi \temp}%
\toksdef\toks@@=1
\def\generalvm@rph#1#2{\ifx#1\rt
    \toks@=\expandafter{\the\toks@
      \rlap{$\vcenter{\rtm@rphtrue\@shiftmorph{#2}}$}}\else
    \toks@@={\llap{$\vcenter{\rtm@rphfalse\@shiftmorph{#2}}$}}%
    \toks@=\expandafter\expandafter\expandafter{\expandafter\the\expandafter
      \toks@@ \the\toks@}\fi \generalvmap}%
\def\m@kevmap{\the\toks@}%
\def\mapdown{\toks@={\vcenter{\drawdownarrow}}\generalvmap}%
\def\mapup{\toks@={\vcenter{\drawuparrow}}\generalvmap}%
\def\bimapdown{\toks@={\vcenter{\drawbidownarrow}}\generalvmap}%
\def\bimapup{\toks@={\vcenter{\drawbiuparrow}}\generalvmap}%
\def\adjmapdown{\toks@={\vcenter{\drawadjdownarrow}}\generalvmap}%
\def\adjmapup{\toks@={\vcenter{\drawadjuparrow}}\generalvmap}%
\def\vline{\toks@={\vcenter{\drawvline}}\generalvmap}%
\def\bivline{\toks@={\vcenter{\drawbivline}}\generalvmap}%
\def\drawdownarrow{\hbox to5pt{\hss\drawvector(0,-1){\varrowlength}\hss}}%
\def\drawuparrow{\hbox to5pt{\hss\drawvector(0,1){\varrowlength}\hss}}%
\def\drawbidownarrow{\hbox to5pt{\hss\hbox{\drawvector(0,-1){\varrowlength}}%
  \hskip\channelwidth\hbox{\drawvector(0,-1){\varrowlength}}\hss}}%
\def\drawbiuparrow{\hbox to5pt{\hss\hbox{\drawvector(0,1){\varrowlength}}%
  \hskip\channelwidth\hbox{\drawvector(0,1){\varrowlength}}\hss}}%
\def\drawadjdownarrow{\hbox to5pt{\hss\hbox{\drawvector(0,-1){\varrowlength}}%
  \hskip\channelwidth\lower\varrowlength
  \hbox{\drawvector(0,1){\varrowlength}}\hss}}%
\def\drawadjuparrow{\hbox to5pt{\hss\hbox{\drawvector(0,1){\varrowlength}}%
  \hskip\channelwidth\raise\varrowlength
  \hbox{\drawvector(0,-1){\varrowlength}}\hss}}%
\def\drawvline{\hbox to5pt{\hss\drawline(0,1){\varrowlength}\hss}}%
\def\drawbivline{\hbox to5pt{\hss\hbox{\drawline(0,1){\varrowlength}}%
  \hskip\channelwidth\hbox{\drawline(0,1){\varrowlength}}\hss}}%
\def\commdiag#1{\null\,
  \vcenter{\commdiagbaselines
  \m@th\ialign{\hfil$##$\hfil&&\hfil$\mkern4mu ##$\hfil\crcr
      \mathstrut\crcr\noalign{\kern-\baselineskip}
      #1\crcr\mathstrut\crcr\noalign{\kern-\baselineskip}}}\,}%
\def\commdiagbaselines{\baselineskip15pt \lineskip3pt \lineskiplimit3pt }%
\def\gridcommdiag#1{\null\,
  \vcenter{\offinterlineskip
  \m@th\ialign{&\vbox to\vgrid{\vss
    \hbox to\hgrid{\hss\smash{$##$}\hss}}\crcr
      \mathstrut\crcr\noalign{\kern-\vgrid}
      #1\crcr\mathstrut\crcr\noalign{\kern-.5\vgrid}}}\,}%
\newdimen\harrowlength \harrowlength=60pt
\newdimen\varrowlength \varrowlength=.618\harrowlength
\newdimen\sarrowlength \sarrowlength=\harrowlength
\newdimen\hmorphposn \hmorphposn=\z@
\newdimen\vmorphposn \vmorphposn=\z@
\newdimen\morphdist  \morphdist=4pt
\dimendef\@hmorphdflt 0
\dimendef\@vmorphdflt 2
\newdimen\hmorphposnrt  \hmorphposnrt=\z@
\newdimen\hmorphposnlft \hmorphposnlft=\z@
\newdimen\vmorphposnrt  \vmorphposnrt=\z@
\newdimen\vmorphposnlft \vmorphposnlft=\z@

\newdimen\hgrid \hgrid=15pt
\newdimen\vgrid \vgrid=15pt
\newdimen\hchannel  \hchannel=0pt
\newdimen\vchannel  \vchannel=0pt
\newdimen\channelwidth \channelwidth=3pt
\dimendef\@hchannel 0
\dimendef\@vchannel 2
\catcode`& = \@oldandcatcode
\catcode`@ = \@oldatcatcode
%
%
%
\def\frac#1#2{{#1\over#2}}

\def\tfrac#1#2{{\textstyle{#1\over#2}}}
%
%
\def\Re{{\Bbb R}}                    
\def\Co{{\Bbb C}}                    
\def\LL{{\Bbb L}}                    
\def\cd{\nabla\ida}                  
\def\d{{\rm d}}                      
\def\de{\partial}                    
\def\dh{\d_{\rm H}}                  
\def\dv{\d_{\rm V}}                  
\def\hor{\mathop{\rm Hor}}           
\def\na{\nabla}                      
\def\rdot{{\raise.5ex \hbox{.}}}     
\def\inn{\mathbin{\rfloor}}          
\def\Lie{\hbox{{\it \$}}}            
\def\Lag{\hbox{\boldsymbols L}}      
\def\*{{\displaystyle*}}             
\def\id{{\rm id}}                    
\def\ida#1#2{\ifx#1^{}^{#2}
             \else\ifx#1_{}_{\!#2}
             \else\errmessage{Sub/Superscript token missing}\fi
             \fi}                    
\def\Im{{\rm Im}}                    
\def\Si{\Sigma}

\def\({\bigl(}
\def\){\bigr)}


\def\rt{\to}

\def\ba{\begin{array}}
\def\ea{\end{array}}

\def\mg{{\bi g}}
\def\bOmega{{\bf\Omega}}
\def\bSigma{{\bf\Sigma}}
\def\bTheta{{\bf\Theta}}
\def\bXi{{\bf\Xi}}
%
%
\hyphenation{Fa-ti-be-ne Frau-en-dien-er Haw-king In-feld La-gran-gian
La-gran-gians
             Mink-ow-ski Pen-rose Reiss-ner Schwarz-schild Waerd-en}
\hbox{ }
\vskip 3truecm
\centre{\baselineskip=1.314\baselineskip\tit
Two-spinor Formulation of First Order Gravity coupled to Dirac Fields}
\vskip 10pt
\centre{by}
\vskip 10pt
\centre{Marco GODINA, Paolo MATTEUCCI,\goodbreak
        Lorenzo FATIBENE \& Mauro FRANCAVIGLIA}
\vskip 10pt
\centre{Dipartimento di Matematica, Universit\`a di Torino\goodbreak
        Via Carlo Alberto 10, 10123 Torino, Italy\goodbreak}

\headline={\sevenrm\ifodd\pageno{}\hfil{\sevenit
Two-spinor Formulation of First Order Gravity coupled to Dirac
Fields\/}\hfil\folio
           \else\folio\hfil M. GODINA, P. MATTEUCCI, L. FATIBENE \& M.
FRANCAVIGLIA
            \hfil{}\fi}
\footline={\sevenrm\hfil\folio\hfil}
%
\baselineskip=15pt

\abstract{Two-spinor formalism for Einstein Lagrangian is
developed. The gravitational field is regarded as a composite object
derived from
soldering forms. Our formalism is geometrically and globally well-defined and
may be used in virtually any $\hbox{{\eightrm 4}{\eighti m}}$-dimensional
manifold
with arbitrary signature as well as without any
stringent topological requirement on space-time, such as parallelizability.
Interactions and feedbacks between gravity and spinor fields are considered.
As is well known, the Hilbert-Einstein Lagrangian is second order also when
expressed in terms of soldering forms.
A covariant splitting is then analysed leading to a first order Lagrangian
which is recognized to play a fundamental role in the theory of conserved
quantities.
The splitting and thence the first order Lagrangian depend on a
reference spin connection which is physically interpreted as setting the
zero level
for conserved quantities. A complete and detailed treatment of conserved
quantities
is then presented.}

\mysec{Introduction}
In the last decade many efforts have been produced in
the literature to provide a better understanding of the new
geometrodynamical variables proposed by Ashtekar [1, 2].
As it is known, Ashtekar's is
a new set of variables for gravity involving soldering
forms and connections. The aim of this paper is to present,
by using only soldering forms as the independent field variables, a covariant
and global first order spinorial splitting of Hilbert's Lagrangian.

A similar splitting was introduced in 1916 by Einstein [3] in order to deal
with the problem of the energy of the gravitational field and, more
generally, with the
problem of conserved quantities associated to the gravitational field itself.
However, Einstein's original splitting was non-covariant and the
conserved quantities so-defined were non-covariant as well.
Later, it was recognized, originally by Rosen in [4] (see also~[5, 6, 7, 8]),
that a covariant splitting was possible, provided that a background connection
is introduced, which then enters the expression of conserved quantities.

Since it is generally accepted that in General Relativity no absolute
quantity should depend on {\it unphysical background\/} fields, one is forced
to interpret these conserved quantities as {\it conserved quantities
relative to the
background\/ {\rm(better, {\it reference\/})} configuration}.
On the other hand, in the literature (see~[9, 10])
it is well accepted that in General Relativity only relative conserved
quantities make sense.
This is intuitively clear if one bears in mind that conserved quantities are
non-local quantities and that solutions in General Relativity may be globally
very different from each other also from a topological viewpoint.
Then it sounds reasonable that, e.g., an infinite amount of energy has to be
spent to deform a solution so much that its global properties change.
In this way, the set of solutions of General Relativity is disconnected into
classes, which are physically separated by an infinite potential barrier.

The starting point of this paper is the observation that the Hilbert
Lagrangian, expressed in spinorial variables, admits
a background-dependent  global and covariant splitting,
in which  the first term is a global formal divergence
playing no role at all for the field equations (since divergences
have vanishing variational derivatives) and the second term
gives a family of first order global Lagrangians,
which generate Einstein's field equations.
The background field, which parametrizes the new family
of global Lagrangians, is a non-dynamical $SL(2,\Co)$ spin connection.
Clearly, the globality of the Lagrangian is useless to ensure the globality of
solutions (general covariance of the equations ensures it),
but plays a fundamental role in the theory of conserved quantities.

Our formalism has been worked out to deal with interactions between
gravity and spinors in a framework which recalls gauge theories in their
geometrical formulation, where one starts from a principal fibre bundle over
space-time, the so-called {\it structure bundle\/} $\Sigma$.
The structure bundle encodes the symmetry structure of the theory.
The {\it configuration bundle\/} $B$ is then associated to the structure
bundle:
i.e., the principal automorphisms of the structure bundle are represented
on $B$ by
means of a natural (functorial) action.

In our formalism, gravity is described by the Ashtekar soldering forms,
which for the first time are here presented as global sections
of a bundle $\Si_\chi$ associated to the structure bundle $\Sigma$.
Globality of soldering forms was already achieved in particular cases
(e.g., on parallelizable manifolds), usually at the cost of requiring
very stringent topological properties on space-time. Our framework applies
to a very wide class of manifolds (namely, to {\it any\/} spin manifold).

The bundle $\Si_\chi$ has been here called the {\it bundle of
(co)spin-vierbeins\/} and, as stated above, is built out of $\Si$
in a canonical (functorial) fashion. These spinorial variables are
suitably related to spin structures on space-time and any co(spin)-vierbein
induces a metric, which is then regarded as a composite object.

In our framework, one does not have to fix the metric $\mg$ on space-time,
give the Lagrangian and thence the field equations (of which $\mg$ has
to be a solution) before defining any spin structure---as on the contrary it
is a standard procedure in the literature when dealing with spinors and
gravity.
Clearly, the standard approach makes sense only when the gravitational field
is considered unaffected by spinors, whereas our formalism is able to describe
the complete interaction and feedback between gravity and spinor fields.

Thus, a field theory for sections of $\Si_\chi$ is considered.
A background $SL(2,\Co)$ spin connection, possibly determined by
a background (co)spin-vierbein, is introduced merely in order to globalize
(in spinorial variables) the local and non-covariant first order Lagrangian
originally  given by Einstein, playing no other role but
setting the ``zero level'' for conserved quantities.

\mynumsec{Spin Structures, spin-frames and soldering forms}
Let $M$ be a (real) 4-dimensional orientable manifold which admits
a smooth metric~$\mg$ of signature $(+,-,-,-)$ and components $(g_{\mu\nu})$;
i.e., we shall assume throughout the sequel that $M$  satisfies the
topological
requirements which ensure the existence on it of a Lorentzian structure
$(M,\mg)$.
We also stress that we are {\it not\/} fixing $\mg$, but it is to be
understood
as determined by the spinorial variables $(e^{AB'}\ida_{\mu})$ giving the
soldering form
as defined below and which will be called ``(co)spin-vierbeins''.

With this end in view, we shall also assume that our space-time~$M$
admits a ``free spin structure'' (see [11, 12, 13, 14] and references
therein);
i.e., we shall assume the existence of at least one principal fibre bundle
$\Sigma$
over $M$ with structure group $SL(2,\Co)$, called the {\it spin structure
bundle},
and at least one strong (i.e.\ covering the identity map) equivariant morphism
$\Lambda:\Sigma\rightarrow \LL(M)$, $\LL(M)$ denoting the principal bundle
of linear frames on $M$. Equivalently, we have the following commutative
diagrams
$$
\commdiag{
\Sigma   &\mapright^{\Lambda} &\LL(M)\cr
\mapdown  &                           &\mapdown\cr
M               &\mapright_{id_{M}}           &M\cr
         }
         \quad \quad \quad
         \commdiag{
\Sigma   &\mapright^{R_{S}} &\Sigma \cr
\mapdown\lft\Lambda&                 &\mapdown\rt\Lambda\cr
\LL(M)               &\mapright_{R_{\hat l(S)}}           &\LL(M)\cr
         }
\eqno (1.1)
$$
where $\hat l:=i\circ l$ is the composed morphism of~$l$,
the epimorphism which exhibits $SL(2,\Co)$  as a two-fold covering of
the proper orthochronous Lorentz group $SO(1,3)_{0}$, with the
canonical injection $i\colon SO(1,3)_{0}\to GL(4)$ of Lie groups,
and $R$ denotes each of the canonical right actions (see~[14]).

We call the bundle map $\Lambda$ a {\it spin-frame\/} on $\Sigma$ and
the pair $(\Sigma,\Lambda)$ a {\it free spin structure}.

This definition of spin structure induces metrics on $M$.
In fact, given a spin-frame $\Lambda:\Sigma\to \LL(M)$,
we can define a metric via the reduced subbundle $SO_{0}(M,\mg_{\Lambda})\equiv
\Im(\Lambda)$ of $\LL(M)$. In other words, $\mg_{\Lambda}$
is the only {\it dynamic\/} metric such that
frames in $\Im(\Lambda)\subset \LL(M)$ are $\mg_{\Lambda}$-orthonormal frames.
It is important here to stress that in our picture the metric $\mg_{\Lambda}$
is built up {\it a posteriori}, after a spin-frame has been determined by the
field equations in a way which is compatible with the (free) spin structure
one has used to define spinors.

This definition of (free) spin structure without fixing
any background metric, which already
appeared in an original work by van den Heuvel~[15],
is given with respect to a fixed spin bundle $\Sigma$, but permitting
variation of spin-frames. The variation of spin-frames
induces a variation of the metric. In fact, it has now been established~[14]
that there is a bijection between spin-frames and sections of
a gauge-natural bundle, here denoted by $\Sigma_{\rho}$,
a fibre bundle the sections of which represent
spin-frames. Such a bundle is given as follows.

Remind that $SL(2,\Co)\cong Spin(1,3)_{0}$ and consider the following
left action of the group $GL(4)\times SL(2,\Co)$ on the manifold
$GL(4)\equiv GL(4,\Re)$
$$
\left\{
\eqalign{
\rho&\colon(GL(4)\times SL(2,\Co))\times GL(4)\to GL(4)\cr
\rho&\colon((A^\mu\ida_\nu,t^{A}\ida_{B}),e^a\ida_\mu)
\mapsto(\Lambda^a\ida_b(\bi t)e^b\ida_\nu(\bi A^{-1})^\nu\ida_\mu)\cr
        }
\right.
\eqno (1.2)
$$
together with the associated  bundle $\Sigma_\rho:=W^{1,0}(\Sigma)
\times_{\rho}GL(4)$, where $W^{1,0}(\Sigma):=\LL(M)\times_{M}\Sigma$
denotes the principal prolongation of order~$(1,0)$ of the principal fibre
bundle~$\Sigma$ and $\times_{M}$ denotes the fibred product of two bundles
over the same base manifold.
The bundle $\LL(M)\times_{M}\Sigma$ is a principal fibre bundle with
structure group $GL(4)\times SL(2,\Co)$.
It turns out that $\Sigma_\rho$ is a fibre bundle associated to
$W^{1,0}(\Sigma)$,
i.e.\ a gauge-natural bundle of order~$(1,0)$. The bundle $\Sigma_\rho$ has
been
called the {\it bundle of (co)spin-tetrads}~[14].

Under these assumptions, to each point $p \in M$
we can assign a complex 2-dimensional vector space $S_{p}(M)$
equipped with a non-degenerate symplectic form (2-form).
(The components of) a generic element of $S_{p}(M)$ will be denoted by
$\xi^A$ and (the components of) the corresponding symplectic form by
$\varepsilon_{AB}$ (its inverse will be denoted by $\varepsilon^{AB}$ and
is such that
$\varepsilon_{AC}\varepsilon^{AB}=\delta_C{}^B$). The complex conjugate vector
space associated with $S_{p}(M)$ will be denoted by $\overline{S_{p}(M)}$,
its elements by  $\bar\xi^{A'}$ and the symplectic form by
$\varepsilon_{A'B'}$.
Since the group preserving the structure on $S_{p}(M)$ is $SL(2,\Co)$,
$\xi^A$ will be called a $SL(2,\Co)$ spinor at $p\in M$ or, for short,
a {\it two-spinor}.

\noindent
Equivalently, two-spinors may be defined via the standard linear action of
$SL(2,\Co)$
on $\Co^{2}$ and we shall denote by $S(M):=(\Sigma\times\Co^{2})/SL(2,\Co)$
the vector bundle associated to the principal fibre bundle $\Sigma$ by
means of this action.
The spin connection can then be used to construct a $SL(2,\Co)$ covariant
derivative of
spinor fields.

Now, if we wish to consider a field theory in which spinorial variables
are dynamical, we must first construct a fibre bundle the sections of which
represent spin-vierbeins. To this end, we need to make a short digression
on complex structures in order to clarify our notation. The material presented
here is standard.

Recall that, if $E$ is a complex vector space, then its
{\it conjugate space\/} $\bar E$ is obtained from $E$ by redefining scalar
multiplication. The new scalar multiplication by $m \in \Co$ is the old scalar
multiplication by $\bar m$. The axioms of a complex vector space are easily
seen
to be satisfied on $\bar E$. Usually, one agrees to denote by $\bar{\bi v}$
the
vector $\bi v$ when it is considered as an element of $\bar E$. If $\bi
f\colon E\to F$
is a linear map of complex vector spaces, then one defines
a linear map $\bar{\bi f}\colon\bar E \to \bar F$ by  $\bar{\bi f}(\bar{\bi
v}):=\overline{\bi f(\bi v)}$. For any complex vector space $E$
the spaces ${(\bar E)}^*:=\{\,\balpha\colon{\bar E}\to\Co\mid\balpha{\rm\
is\ linear}\,\}$ and $\overline{(E^*)}:=\{\,\bar\bbeta\mid\bbeta\colon
E\to\Co{\rm\ is\ linear}\,\}$ are naturally isomorphic. The isomorphism
$\biota\colon
{(\bar E)}^* \to\overline{(E^*)}$ is given by
$\biota(\balpha):={\bar\bbeta}$, where
$\langle\bbeta,\bi v\rangle=\overline{\langle\balpha,\bar{\bi v}\rangle}$ and
$\bi v \in E$. Owing to such an isomorphism, we shall identify the space
${(\bar E)}^*$
with $\overline{(E^*)}$ and denote it ${\bar E}^*$.
Let us also recall that, in general, for a complex vector space $E$ there
is no canonical way to represent $E$ as the direct sum of two real spaces,
the {\it real\/} and {\it imaginary\/} parts of $E$, although each complex
vector space $E$ admits a {\it real form\/} obtained by taking
the same set and restricting the scalars to be real. An additional {\it
real structure\/} in $E$ (see, e.g., [16]) is a linear map $C\colon
E\to\bar E$ such that $\bar CC=\id_{E}$. Any vector
$\bi v\in E$ splits as $\bi v=\bi v^{+}+\bi v^{-}$, where we
set $\bi v^{\pm}:={1\over2}(\bi v\pm\bar C\bar{\bi v})$.
We have a direct sum decomposition of $E$ into two real vector spaces
$E^{+}$ and $E^{-}$ such that $\bi v \in E^{\pm}$ iff $\bar{\bi v}=\pm C\bi v$.
On the vector space $E=\Co^{2}\otimes\bar{\Co}^{2}$ consider the real structure
$C\colon\Co^{2}\otimes{\bar\Co^{2}}\to\bar\Co^{2}\otimes\Co^{2}$ defined by
$C(\bi u\otimes\bar{\bi v}):=\bar{\bi v}\otimes\bi u$. The real space
$E^{+}$ is the real space of Hermitian tensors spanned by elements of the
form $\bi u\otimes\bar{\bi v}$. A generic element
of $E^{+}$ is written as $\bphi=\phi^{AB'}{\bi c}_{A}\otimes{\bi c}_{B'}$
where
$\overline{\phi^{AB'}}=\phi^{BA'}$ and $({\bi c}_{A'})$ is the basis
of $\bar{\Co}^{2}$ consisting of the same vectors as $({\bi c}_{A})$.
Hermitian tensors of the real vector space $E^{+}$ are also called {\it real\/}
(see Ref.~[17]).

\noindent
Now, let $V$ be the open subset of
$E^{+}\otimes(\Re^{4})^{*}$ consisting of all invertible real linear maps
$\bphi\colon\Re^{4}\to E^{+}$.
An element $\bphi\equiv\phi^{AB'}\ida_{\mu}{\bi c}_{A}\otimes{\bi
c}_{B'}\otimes{\bi c}^{\mu}$
of the vector space $\Co^{2}\otimes{\bar\Co^{2}}\otimes(\Re^{4})^{*}$
belongs to $V$ iff the following conditions hold
$$
\displaylines{
\hfill
\overline{\phi^{AB'}\ida_{\mu}}=\phi^{BA'}\ida_{\mu},
\hfill\llap{$(1.3a)$}\cr\hfill
\phi^{AB'}\ida_{\mu}\phi_{AB'}\ida^{\nu}=\delta^\nu_\mu,
\hfill\llap{$(1.3b)$}\cr\hfill
\phi^{AB'}\ida_{\mu}\phi_{CD'}\ida^{\mu}=\delta^{A}_{C}\delta^{B'}_{D'},
\hfill\llap{$(1.3c)$}\cr
          }
$$
where $(\phi_{AB'}\ida^{\mu})$ denote the components of the inverse element
$\bphi^{-1}\equiv\phi_{AB'}\ida ^{\mu}{\bi c}^{A}\otimes{\bi
c}^{B'}\otimes{\bi c}_{\mu}$; here indices are {\it not\/} raised or lowered
with $g_{\mu\nu}$ or $\varepsilon_{AB}$, although, if we define $g_{\mu\nu}:=
\phi^{AB'}\ida_{\mu}\phi^{CD'}\ida_{\nu}\varepsilon_{AC}\varepsilon_{B'D'}$, the
n we find
$\phi_{BA'}\ida^{\mu}=\phi^{\>\>\cdot\>\>\cdot}_{BA'}\ida^{\mu}_{\cdot}$,
where on the r.h.s.\ the tensor index~$\mu$ is raised using $g_{\mu\nu}$
and the
indices~$AB'$ are lowered using $\varepsilon_{AB}$ and
$\varepsilon_{A'B'}$, respectively. Formulae~$(1.3b)$ and~$(1.3c)$ reflect
the fact that the composed linear map $\bphi^{-1}\circ\bphi$ is the
identity map on $\Re^{4}$ and $\bphi\circ\bphi^{-1}$ is the identity map
on~$E^{+}$.

We are at last in a position to consider the following left action on $V$
$$
\left\{
\eqalign{
\chi&\colon(GL(4)\times SL(2,\Co))\times V\to V\cr
\chi&\colon((A^{\mu}\ida_{\nu},t^{A}\ida_{B}), W^{AB'}\ida_{\nu})
\mapsto(t^{A}\ida_{C}t^{B'}\ida_{D'}
        W^{CD'}\ida_{\nu}(\bi A^{-1})^{\nu}\ida_{\mu})\cr
        }
\right.
\eqno (1.4)
$$
together with the associated  bundle
$\Sigma_\chi:=(\LL(M)\times_{M}\Sigma)\times_{\chi}V$.
According to the theory of gauge-natural bundles and gauge-natural operators
(see Ref.~[18]), $\Sigma_\chi$ turns out to be a fibre bundle associated to
$W^{1,0}(\Sigma)$, i.e.\ a gauge-natural bundle of order $(1,0)$.
Local coordinates on the bundle $\Sigma_\chi$ will be denoted by
$(x^{\mu}, e^{AB'}\ida_{\mu})$. A section of $\Sigma_\chi$
will be called a {\it (co)spin-vierbein}.
Equivalently, a (co)spin-vierbein may be regarded as an Ashtekar
soldering form, i.e.\ as an (invertible) linear map $\bi A_{p}\colon T_{p}M
\to{S_{p}(M)}\otimes\overline{S_{p}(M)}$ at each point $p\in M$, with the
property of being ``real'', i.e.\ such that the components
$(A^{AB'}\ida_{\mu})$
of $\bi A_{p}$ constitute a Hermitian matrix for each value of $\mu$.

It is possible to construct another bundle $\Sigma_\tau$ with the same
fibre $V$ by considering the following left action on the $SL(2,\Co)$-%
manifold~$V$:
$$
\left\{
\eqalign{
\tau&\colon SL(2,\Co)\times V\to V\cr
\tau&\colon(t^{A}\ida_{B}, F^{AB'}\ida_{a})
\mapsto(t^{A}\ida_{C}t^{B'}\ida_{D'}
        F^{CD'}\ida_{b}\Lambda^{b}\ida_{a}(\bi t^{-1}))\cr
        }
\right..
\eqno (1.5)
$$
The bundle $\Sigma_\tau:=\Sigma\times _{\tau}V$ is a fibre bundle
associated to the principal fibre bundle $\Sigma$,
also denoted by $W^{0}(\Sigma)$, with structure group $SL(2,\Co)$.
It turns out that $\Sigma_\tau$ is a gauge-natural bundle of order zero,
i.e.\ associated to the ``trivial'' (zeroth order) principal prolongation
of $\Sigma$. Local coordinates on the bundle $\Sigma_\tau$ will be denoted by
$(x^{\mu},M^{AB'}\ida_{a})$. A special choice for $\Sigma_\tau$ is the section
$\sigma_{{\rm IW}}\colon M\to\Sigma_\tau$ whose components,
in any system of local coordinates,
are given by the ``Infeld-van der Waerden symbols'' [11, 19--22], i.e.\ the
section
$$
\sigma_{{\rm IW}}\colon(x^\alpha)
\mapsto(x^\alpha,M^{AB'}\ida_{a}=\sigma^{AB'}\ida_{a}).
\eqno{(1.6)}
$$
The {\it Infeld-van der Waerden section\/} $\sigma_{{\rm IW}}$ shall be called
the {\it canonical section\/} of $\Sigma_\tau$. It is a global section because
its components, i.e.\ the Infeld-van der Waerden symbols, are the
components (in the
standard fibre) of an $SL(2,\Co)$-invariant tensor.

In fact, whenever one has a principal fibre bundle $(P,M,G,\pi)$ with
structure group~$G$ and a left action of~$G$ on some real or complex
vector space~$V$, it is possible, if we are given an invariant vector
of~$V$ with respect to~$G$, i.e.\ if we suppose there exists a
vector $\bi v\in V$ such that $g\cdot\bi v=\bi v$ for all $g\in G$, to
construct (using the transition functions of~$P$) a global section~$s$
of the associated bundle~$(P\times V)/G$, whose components are the components
of $\bi v\in V$ with respect to a basis chosen in~$V$.

The {\it canonical\/} Infeld-van der Waerden section
induces a {\it canonical isomorphism\/} (over the identity) of
real fibre bundles, locally represented by:
$$
\left\{
\eqalign{
\Phi_{{\rm IW}}&\colon\Sigma_\chi\to\Sigma_\rho\cr
\Phi_{{\rm IW}}&\colon(x^\alpha, e^{AB'}\ida_{\mu})
\mapsto (x^\alpha,e^{a}\ida_{\mu}=\sigma_{AB'}\ida^{a}e^{AB'}\ida_{\mu})\cr
        }
\right..
\eqno (1.7)
$$

We are in a position to state the following (cf.\ Ref.~[14])

\noindent {\bf Proposition.}
{\it There is a bijection between spin-frames and the sections of
the gauge-natural bundle $\Sigma_{\chi}$, i.e.\ between spin-frames and
(co)spin-vierbeins (Ashtekar soldering forms).}

In other words, the above proposition asserts that we may
represent spin-frames with {\it dynamical\/} (global) Ashtekar soldering
forms,
and this fact is crucial if we want to consider a field theory
in which spin-frames are dynamical.

\mynumsec{Standard General Relativity in two-spinor formalism}
In our theory the standard ``Hilbert'' spinor Lagrangian is built out of
the soldering form variables, or our ``(co)spin-vierbeins''. Of course,
it turns out to be a second order Lagrangian theory in these variables.

In fact, define
$$
\bSigma^{AB}:=\frac{i}{2}\varepsilon_{A'B'}\,\btheta^{AA'}\wedge
                  \btheta^{BB'},
\eqno (2.1)
$$
$\btheta^{AA'}=e^{AA'}\ida_{\mu}\,\d x^{\mu}$ being the
{\it Ashtekar soldering form\/} [1, 2, 23]. Define also
$$
\bOmega^{A}\ida_{B}:=\dh\bomega^{A}\ida_{B}+\bomega^{A}\ida_{C}\wedge
                     \bomega^{C}\ida_{B},
$$
where $\dh$ is the horizontal differential~[24] and the coefficients of the
(unprimed) spin connection $\bomega^{A}\ida_{B}\equiv\omega^{A}\ida_{B\mu}\,
\d x^{\mu}$ are regarded as being uniquely determined by the spin-vierbeins
and their first partial derivatives $(e^{AB'}\ida_{\mu\nu})$ via the relation
(cf.~[25, 26])
$$
\omega^{A}\ida_{B\mu}={1\over 2}(e_{BA'}\ida^\nu e^{AA'}\ida_{[\mu\nu]}+
                      e^{AA'\rho}e_{CC'\mu}e_{BA'}\ida^\nu
                      e^{CC'}\ida_{[\rho\nu]}+e^{AA'\nu}e_{BA'[\nu\mu]}).
\eqno{(2.2)}
$$
Since we aim to describe a spinor field (without any further gauge
symmetry) in interaction with gravity, our {\it configuration space\/}
will be assumed to be the following bundle
$$
B=\Sigma_{\chi}\times_{M}\Sigma_{\gamma},
\eqno{(2.3)}
$$
where $\Sigma_{\gamma}:=\Sigma\times_{\gamma}E$ is the vector bundle
associated to the principal bundle~$\Sigma$ via the obvious representation~%
$\gamma$ of $SL(2,\Co)$ on the vector space~$E:=\bar\Co^2\oplus(\Co^2)^*$.
The bundle~$\Sigma_{\gamma}$ is then isomorphic to $\bar S(M)\oplus_M
S^*(M)$.

\noindent
Consequently, the Lagrangian will be chosen of the following form:
$$
\Lag\colon J^{2}\Sigma_{\chi}\times_{M}J^{1}\Sigma_{\gamma}
    \to\Lambda^{4}T^*\!M.
\eqno{(2.4)}
$$
According to the principle of minimal coupling, the Lagrangian~$\Lag$
is assumed to split into two parts $\Lag=\Lag_{H}+\Lag_{D}$, where
$$
\left\{
\eqalign{
\Lag_H &\colon J^{2}\Sigma_{\chi}\to\Lambda^4T^*\!M\cr
\Lag_H &=-{1\over\kappa}{\bf \Omega}_{AB}\wedge{\bf\Sigma}^{AB}+{\rm
c.c.}\qquad
         (\kappa:=8\pi G/c^{4})
}
\right.
\eqno (2.5)
$$
is the standard ``Hilbert'' spinor Lagrangian, ``c.c.'' stands for the
complex conjugate of the preceding term
and $\Lag_{D}\colon J^{1}(\Sigma_{\chi}\times_{M}
\Sigma_{\gamma})\to\Lambda^4T^*\!M$ is the two-spinor equivalent
of the Dirac Lagrangian [14, 26]
$$
\bi L_{D}=\left[{i\over2}(\tilde{\bold\Psi}\cdot\gamma^{a}\cdot\cd_{a}
\bold\Psi-\widetilde{\cd_{a}\bold\Psi}\cdot\gamma^{a}\cdot\bold\Psi)
-m\tilde{\bold\Psi}\cdot\bold\Psi\right]\bSigma,
$$
where $\tilde{\bold\Psi}:=\bold\Psi^{\dagger}\cdot\gamma_{0}$
is called the {\it Dirac adjoint\/} of~$\bold\Psi$,
$\gamma^{a}:=\eta^{ab}\gamma_{b}$, the dot `$\cdot$' denotes matrix product
and $\bSigma:={^4}e\,{\bf ds}$ is the standard $4$-form,
${^4}e$ being the determinant of $(e^{a}\ida_{\mu})$
(or, equivalently, the determinant of $(e^{AB'}\ida_{\mu})$)
and ${\bf ds}:=\d x^0\wedge\d x^1\wedge\d x^2\wedge\d x^3$
the (local) volume element.
If we set $\bold\Psi=:\bpsi\oplus\bvarphi$, $\tilde{\bold\Psi}=:\bar\bvarphi
\oplus\bar\bpsi$ (see Ref.~[17]) and define for any vector field $\bi v$
on~$M$
$$
\not{\!v}\bold\Psi\equiv v^{a}\gamma_{a}\cdot\bold\Psi:=
     \sqrt{2}(v^{AA'}\varphi_{A}\bi f_{A'}\oplus v_{AA'}\psi^{A'}\bi f^{A})
$$
(which implies our {\it Clifford product\/} has the form
$\gamma_{a}\cdot\gamma_{b}+\gamma_{b}\cdot\gamma_{a}=\eta_{ab}\Bbb I_{4}$),
$\bpsi\equiv\psi^{A'}\bi f_{A'}$ being a section of~${\bar S}(M)$ and
$\bvarphi\equiv\varphi_{A}\bi f^{A}$ a section of~$S^{*}(M)$, it is
straightforward to prove that~$\Lag_{D}$ has the following expression
$$
\Lag_{D}=\biggl\{\biggl[{i\sqrt{2}\over 2}(\bar\varphi_{A'}\cd^{AA'}\varphi_{A}
         +\bar\psi^{A}\cd_{AA'}\psi^{A'})
         -m\,\varphi_{A}\bar\psi^{A}\biggr]+{\rm c.c.}\biggr\}\bSigma,
\eqno (2.6)
$$
where $\cd_{AA'}:=e_{AA'}\ida^{\mu}\cd_{\mu}$.

Notice that in this formalism Dirac's equation
$(i\,/\!\!\!\!\na-m)\bold\Psi=\bold 0$ takes the symmetric form
$$
\left\{
\eqalign{
i\sqrt{2}\cd_{AA'}\psi^{A'}-m\varphi_{A} &=0\cr
i\sqrt{2}\cd^{AA'}\varphi_{A}-m\psi^{A'} &=0
}
\right..
\eqno (2.7)
$$
Thus the total Lagrangian $\Lag$ can be simply represented in terms of the
variables discussed above together with their partial derivatives up to
the second order included $(e^{AB'}\ida_{\mu},e^{AB'}\ida_{\mu\nu},
e^{AB'}\ida_{\mu\nu\rho})$.

According to recent results~[27], to each higher order Lagrangian there
corresponds at least one global {\it Poincar\'e-Cartan form}. Such a
form is unique for first order theories; in the second order case
uniqueness is lost, although there is still a canonical choice, which we
will now describe. Let
$$
\bi L\equiv L(x^{\alpha};y^{\bi a},y^{\bi a}\ida_{\lambda},
y^{\bi a}\ida_{\lambda\mu})\,{\bf ds}
$$
be a second order Lagrangian, where $y^{\bi a}$ is a field of
arbitrary nature. Define the {\it momenta\/} by setting
$$
f_{\bi a}\ida^{\lambda\mu}:=\frac{\de L}{\de y^{\bi a}\ida_{\lambda\mu}},
\qquad
f_{\bi a}\ida^{\lambda}:=\frac{\de L}{\de y^{\bi a}\ida_{\lambda}}-
                       \d_{\mu}\frac{\de L}{\de y^{\bi a}\ida_{\lambda\mu}},
$$
where $\d_{\mu}$ denotes the formal derivative [14, 24]. The Poincar\'e-Cartan
form associated to~$\bi L$ is thence given by
$$
\bTheta(\bi L):=\bi L+(f_{\bi a}\ida^{\lambda}\,\dv y^{\bi a}+
                       f_{\bi a}\ida^{\lambda\mu}\,\dv y^{\bi a}\ida_{\mu})
                      \wedge{\bf ds}_{\lambda},
\eqno (2.8)
$$
where $\dv$ is the vertical differential~[24] and we set ${\bf ds}_{\lambda}:=
\de_{\lambda}\inn{\bf ds}$, `$\inn$' denoting inner product.

\noindent
The knowledge of the Poincar\'e-Cartan form enables us to calculate the
so-called {\it energy density flow\/} of the Lagrangian in question. In
fact, if $\bi L$ is a Lagrangian defined on the $k$-th order
prolongation of a gauge-natural bundle~$B$ (see Ref.~[18]) and $\bXi$ is
the generator of a one-parameter subgroup of automorphisms of~$B$, the
energy density flow associated to~$\bi L$ along the vector
field~$\bXi$ is given by (cf.\ [28, 8])
$$
\bi E(\bi L,\bXi)\equiv E^{\alpha}(\bi L,\bXi)\,{\bf ds}_{\alpha}:=
                        -\hor[\tilde\bXi\inn\bTheta(\bi L)],
$$
where $\hor$ denotes the horizontal operator on forms [24] and $\tilde\bXi$
is the $(2k-1)$-th order prolongation of~$\bXi$
(we stress that the word ``energy'' is used here in the broader sense
of ``conserved N\"other current'').
In particular, for a second order Lagrangian one finds:
$$
\bi E(\bi L,\bXi)=(f_{\bi a}\ida^{\alpha}\Lie_{\!\bXi}y^{\bi a}+
                   f_{\bi a}\ida^{\alpha\mu}\Lie_{\!\bXi}y^{\bi a}\ida_{\mu}
                   -L\xi^{\alpha})\,{\bf ds}_{\alpha},
\eqno (2.9)
$$
$\bxi$ being the projection of~$\bXi$ on~$M$.

\noindent Our Poincar\'e-Cartan form associated to $\Lag_{H}$ is
$$
\bTheta(\Lag_{H})=\Lag_{H}-\frac{1}{\kappa}({\bi V}_{AB}\wedge
                  \bSigma^{AB}+{\rm c.c.}),
$$
where ${\bi V}_{AB}:=1/2\,e_{A}\ida^{A'}\ida_{\alpha}e_{BA'}\ida^{\beta}
\,\dv\Gamma^{\alpha}\ida_{\beta\mu}\wedge\d x^{\mu}$ and
$\Gamma^{\alpha}\ida_{\beta\mu}$ is the Levi-Civita connection
induced by the metric $g_{\mu\nu}$, uniquely and unequivocally
determined by the soldering form via the relation
$$
g_{\mu\nu}=e^{AB'}\ida_{\mu}e^{CD'}\ida_{\nu}\varepsilon_{AC}\varepsilon_{B'D'}.
$$
An easy calculation shows that the expression for $\bi E(\Lag_{H})$ is
$$
\bi E(\Lag_{H},\bXi)=-\frac{1}{\kappa}G^{\alpha}\ida_{\beta}\xi^{\beta}
                      \bSigma_{\alpha}
                     +\frac{1}{2\kappa}\dh(\cd_{A}^{A'}\xi_{BA'}\bSigma^{AB}
                     +{\rm c.c.}),
\eqno (2.10)
$$
where $G^{\alpha}\ida_{\beta}$ is the Einstein tensor,
$\bSigma_{\alpha}:=\de_{\alpha}\inn\bSigma$ and we set
$\xi^{AA'}:=e^{AA'}\ida_{\mu}\xi^{\mu}$, $(\xi^{\mu})$ being the
components of~$\bxi$ in a local chart.

\noindent
As one can see from~(2.10), $\bi E(\Lag_{H},\bXi)$ is conserved in
vacuum along any solution of the field equations $G_{\mu\nu}=0$, while
of course it is not in interaction with matter. The 2-form
$$
\bi U(\Lag_{H},\bXi):=\frac{1}{2\kappa}\cd_{A}^{A'}\xi_{BA'}\bSigma^{AB}
                      +{\rm c.c.}
\eqno (2.11)
$$
is called the {\it Hilbert superpotential\/}: it is straightforward to
show that it is nothing but the half of the well known Komar
superpotential~[29]. Therefore, setting (in spherical coordinates)
$\bxi=\de/\de t$ and integrating~(2.11) on a spherical surface, it will
yield half the mass for the Schwarzschild solution, but the correct angular
momentum for the (Schwarzschild and) Kerr solution (see Ref.~[8]).

We can now tackle the spinorial contribution, writing down the
Poincar\'e-Cartan form associated to~$\Lag_{D}$. Using~(2.8), which
is of course still valid for first order Lagrangians as a trivial
subcase, we find:
$$
\eqalign{
&\bTheta(\Lag_{D})=\Lag_{D}+\Bigl\{\tfrac{i\sqrt{2}}{2}[\bar\varphi_{A'}
e^{BA'\alpha}\,\dv\varphi_{B}+\bar\psi^{A}
e_{AB'}\ida^{\alpha}\,\dv\psi^{B'}\cr
&\quad{-\tfrac{1}{2}}(\bar\varphi_{A'}\varphi_{B}
e^{B}\ida_{C'}\ida^{\mu}e_{C}\ida^{A'\alpha}+\bar\psi^{A}\psi^{B'}
e_{CB'}\ida^{\mu}e_{AC'}\ida^{\alpha})\,\dv e^{CC'}\ida_{\mu}]+{\rm
c.c.}\Bigr\}\wedge
\bSigma_{\alpha}.\cr
          }
\eqno (2.12)
$$
Resorting as usual to~(2.9) and making use of the relation (cf.~[14])

$$
\Lie_{\!\bXi}e^{AA'}\ida_{\mu}=\cd_{\mu}\xi^{\nu}e^{AA'}\ida_{\nu}
             -e^{BA'}\ida_{\mu}{\rm V}\Xi^{A}\ida_{B}
             -e^{AB'}\ida_{\mu}\overline{{\rm V}\Xi}{}^{A'}\ida_{B'},
             \eqno (2.13)
$$
where ${\rm V}\Xi^{A}\ida_{B}$ is the vertical part of $\Xi^{A}\ida_{B}$
with respect to the dynamical connection $\omega^{A}\ida_{B\mu}$,
i.e.\ ${\rm V}\Xi^{A}\ida_{B}:=\Xi^{A}\ida_{B}-
\omega^{A}\ida_{B\mu}\xi^{\mu}$, $(\xi^{\mu},\Xi^{A}\ida_{B})$ obviously
being the components of~$\bXi$ in a local chart, we finally get
$$
\bi E(\Lag_{D},\bXi)=T^{\alpha}\ida_{\beta}\xi^{\beta}
                      \bSigma_{\alpha}
                     +\dh\bi U(\Lag_{D},\bXi),
\eqno (2.14)
$$
where $T^{\alpha}\ida_{\beta}$ is the energy-momentum tensor associated
to~$\Lag_{D}$ and we set
$$
\bi U(\Lag_{D},\bXi):=\frac{i\sqrt{2}}{4}\xi_{A}^{A'}
(\bar\varphi_{A'}\varphi_{B}-\bar\psi_{B}\psi_{A'})\bSigma^{AB}
                      +{\rm c.c.}
\eqno (2.15)
$$
Thus, the {\it total energy density flow}
$$
\bi E(\Lag,\bXi)\equiv\bi E(\Lag_{H},\bXi)+\bi E(\Lag_{D},\bXi)
$$
appears to be conserved ``on shell'' (i.e.\ along any solution of the
field equations), owing to the Einstein equations $G^{\alpha}\ida_{\beta}=
\kappa T^{\alpha}\ida_{\beta}$ and in accordance with the general
theory~[28]. As a consequence, the 2-form
$$
\bi U(\Lag,\bXi)\equiv\bi U(\Lag_{H},\bXi)+\bi U(\Lag_{D},\bXi)
$$
can be uniquely identified as the {\it (total) superpotential\/} of the
theory. Notice that the vertical contribution, i.e.\ the one containing
${\rm V}\bXi$, vanishes identically off shell in~(2.14). So, even though
the interpretation of our conserved currents could result difficult in
principle as we enlarged the symmetry group by adding the vertical
transformations, we find out that actually this is not the case.

\mynumsec{Global first order spinor Einstein Lagrangians}

In the theory we developed, we chose as the gravitational part of our
Lagrangian~$\Lag$ the usual ``Hilbert" spinor Lagrangian~(2.5). Another
possible candidate is the back\-ground-dependent family of
{\it global first order\/} Lagrangians
$$
\left\{
\eqalign{
\Lag_G&\colon J^{1}\Sigma_{\chi}\to\Lambda^4T^*\!M\cr
\Lag_G&:=-{1\over\kappa}({\bf K}_{AB}+\bi Q^{C}\ida_{A}\wedge\bi Q_{BC}
                         )
                        \wedge\bSigma^{AB}
         +{\rm c.c.}
}
\right.,
\eqno (3.1)
$$
where ${\bf K}_{AB}$ is the curvature $2$-form of a
background spin connection $\bbeta_{AB}$ (see Ref.~[30] for the basic
formalism)
and we set $\bi Q_{AB}:=\bomega_{AB}-\bbeta_{AB}$.

\noindent If we take $\bbeta_{AB}\equiv\bf 0$ in~(3.1), we recover
but the {\it local\/} non-covariant first order {\it spinor Einstein
Lagrangian\/} of M{\o}ller-Nester~[31, 32];
see also Refs.~[33] and~[34].

We shall call the Lagrangian (3.1), in a given background,
the {\it global first order spinor Einstein Lagrangian}.

Again, relying on the pull-back properties of Poincar\'e-Cartan
forms~[24], we find
$$
\bTheta(\Lag_{G})=\Lag_{G}+\frac{1}{\kappa}[(\dv\bSigma^{AB}+\bSigma^{AC}
                    \wedge{\bi Z}^{B}\ida_{C}+\bSigma^{BC}\wedge
                    {\bi Z}^{A}\ida_{C})\wedge\bi Q_{AB}+{\rm c.c.}],
\eqno (3.2)
$$
where we set ${\bi Z}^{A}\ida_{B}:=1/2\,e^{AA'}\ida_{\mu}\,
\dv e_{BA'}\ida^{\mu}$.

Now, using~(2.9), we can calculate the energy density flow, which,
after some manipulations, appears to be
$$
\bi E(\Lag_{G},\bXi)=-\frac{1}{\kappa}G^{\alpha}\ida_{\beta}\xi^{\beta}
                        \bSigma_{\alpha}+\dh\bi U(\Lag_{G},\bXi)
                        +\frac{1}{2\kappa}(e_{B}\ida^{A'}\ida_{\alpha}
                        e_{AA'}\ida^{\beta}\Lie_{\!\bxi}
                        {\rm B}^{\alpha}\ida_{\beta\mu}\,\d x^{\mu}\wedge
                        \bSigma^{AB}+{\rm c.c.})
\eqno (3.3)
$$
with
$$
\bi U(\Lag_{G},\bXi):=\bi U(\Lag_{H},\bXi)+\frac{1}{\kappa}\,\bxi\inn
                      (\bi Q_{AB}\wedge\bSigma^{AB}+{\rm c.c.}).
\eqno (3.4)
$$
Here, as for the background linear connection ${\rm
B}^{\alpha}\ida_{\beta\mu}$,
we have two possible choices: if ${\rm B}^{A}\ida_{B\mu}$ is given---{\it
mutatis
mutandis\/}---by formula $(2.2)$ via a background soldering form
$f^{AB'}\ida_{\mu}$,
then ${\rm B}^{\alpha}\ida_{\beta\mu}$ is taken to be the Levi-Civita
connection
of the ``{\it induced\/} background metric''
$h_{\mu\nu}:=f^{AB'}\ida_{\mu}f^{CD'}\ida_{\nu}
\varepsilon_{AC}\varepsilon_{B'D'}$;
otherwise, i.e.\ if ${\rm B}^{A}\ida_{B\mu}$ is a
{\it generic\/} background $SL(2,\Co)$ connection and $f^{AB'}\ida_{\mu}$ a
further
background soldering form, ${\rm B}^{\alpha}\ida_{\beta\mu}$ is taken to be
the linear
connection (with torsion) given by the following formula:
$$
{\rm B}^{\alpha}\ida_{\beta\mu}= f_{AB'}\ida^{\alpha}{\rm
B}^{A}\ida_{C\mu}f^{CB'}\ida_{\beta}
+f_{BA'}\ida^{\alpha}{\bar{\rm B}}^{A'}\ida_{C'\mu}f^{BC'}\ida_{\beta}
+f_{AB'}\ida^{\alpha}f^{AB'}\ida_{\beta\mu}.
\eqno (3.5)
$$
Let us note that, considering the soldering form $f^{AB'}\ida_{\mu}$
more generally as an object on a $GL(2,\Co)$-principal bundle $P$
(\dots they are not spinors!), formula $(3.5)$ gives us
a bijection between $GL(2,\Co)$-principal connections on $P$ and
(complex) linear connections on $M$ (i.e.\ complex linear connections on
the complexified tangent bundle $(TM)^{\Co}$ over the real manifold $M$).
These objects are classical and appear in the ``old''
literature (see Refs.~[20, 21, 25, 35]).
That is why, in this case, we shall call them
the {\it Infeld-van der Waerden variables}.

Now, comparing (3.3) with (2.10), we see that in~(3.3) we have two
additional terms containing the Lie derivative of the background
connection. So $\bi E(\Lag_{G},\bXi)$ will be conserved on shell only
for those vector fields~$\bxi$ such that
$\Lie_{\!\bxi}{\rm B}^{\alpha}\ida_{\beta\mu}=0$, e.g.\ for Killing
vector fields of the background linear connection.
It is possible to show~[8] that the additional background contribution
in~(3.4),
when integrated on a spherical surface with $\bxi=\de/\de t$, restores the
expected
value for the mass of the Schwarzschild solution, if the Levi-Civita
connection of
the Minkowski metric is chosen as the (obviously) appropriate background.
Of course,
the angular momentum associated to the Schwarzschild and Kerr solutions is
unaffected
by the background contribution.

There are other good reasons why one should be interested in working
with the new Lagrangian~(3.1) rather than with the usual Hilbert Lagrangian.
In fact, the superpotential $\bi U(\Lag_{G},\bXi)$
reproduces the usual ADM mass~[36] for
asymptotically flat space-times~[37]. Moreover, in the Reissner-Nordstr\"om
case it recovers Penrose's quasi-local mass~[38] (cf.~[39]): when the
global Lagrangian is considered (i.e.\ when also the electrostatic
contribution is taken into account), the result we get is---in our
opinion---even more convincing from a physical point of view~[37].

\noindent
Actually, in the Schwarzschild and Reissner-Nordstr\"om cases, the mass
scalar deriving from $\bi U(\Lag_{G},\bXi)$ coincides with a well-known
definition of mass for spherically symmetric space-times (see~[40] and
references quoted therein). This coincidence is limited to a very
restricted subclass of solutions, although, e.g., both methods
consistently yield the same result for the total mass of a closed FRW
universe, i.e.\ zero~[41].


\mysec{Acknowledgements}
One of us (M.G.) would like to express his gratitude to his twin brother Paolo.

\noindent
P.M. wishes to dedicate this paper to the memory of his beloved godfather
Norman Osborne.

\mysec{References}
{\parindent=20pt\sfcode`\.=1000\tolerance=9999

\bibitem
A. Ashtekar, G.T. Horowitz \& A. Magnon (1982), {\it Gen. Rel. Grav.\/}
{\bf 14},
411--428.

\bibitem
A. Ashtekar (1986), `New variables for classical and quantum gravity',
{\it Phys. Rev. Lett.\/} {\bf 57},
2244--2247.

\bibitem
A. Einstein (1916), {\it Sitzungsberg. Preu{\ss}. Akad. Wiss.\/} (Berlin),
1111;
{\it id.} (1916), {\it Ann. Phys.} {\bf 49}, 769.

\bibitem
N. Rosen (1940), {\it Phys. Rev.} {\bf 57}, 147.

\bibitem
L. Rosenfeld (1940), `Sur le tenseur d'impulsion-energie',
{\it Mem. Roy. Acad. Belg. Cl. Sci.} {\bf 18} No. 6, pp. 1--30.

\bibitem
R. Sorkin (1977), {\it Gen. Rel. Grav.\/} {\bf 8}, 437.

\bibitem
J. Katz (1985), {\it Class. Quant. Grav.\/} {\bf 2}, 423.

\bibitem
M. Ferraris \& M. Francaviglia (1990), {\it Gen. Rel. Grav.\/} {\bf 22} (9),
965--985.

\bibitem
J.D. Brown \& J.W. York (1993), {\it Phys. Rev.\/} D{\bf 47} (4), 1407.

\bibitem
S.W. Hawking \& C.J. Hunter (1998), {\it hep-th/9808085\/};
C.J. Hunter (1998), {\it gr-qc/9807010\/};
S.W. Hawking, C.J. Hunter \& D.N. Page, {\it hep-th/9809035\/}.

\bibitem
R. Penrose \& W. Rindler (1984), {\it Spinors and space-time}, vol.~1,
Cambridge University Press, Cambridge.

\bibitem
A. Haefliger (1956), {\it Comptes Rendus Acad. Sc. Paris\/} {\bf 243},
558--560.

\bibitem
J. Milnor (1963), {\it Enseignement Math.\/} {\bf 9} (2), 198--203.

\bibitem
L. Fatibene, M. Ferraris, M. Francaviglia \& M. Godina (1998),
{\it Gen. Rel. Grav.\/} {\bf 30} (9), 1371--1389.

\bibitem B.M. van den Heuvel (1994),
{\it J. Math. Phys.} {\bf 35} (4), 1668--1687.

\bibitem
P. Budinich \& A. Trautman (1988), {\it The Spinorial Chessboard},
Springer-Verlag, New York.

\bibitem
R.S. Ward \& R.O. Wells Jr. (1990), {\it Twistor Geometry and Field
Theory}, Cambridge University Press, Cambridge.

\bibitem
I. Kol{\'a}{\v r}, P.W. Michor \& J. Slov\'ak (1993), {\it Natural
Operations in Differential Geometry}, Springer-Verlag, Berlin.

\bibitem
F. de Felice \& C.J.S. Clarke (1990), {\it Relativity on curved manifolds},
Cambridge University Press, Cambridge.

\bibitem
L. Infeld \& B.L. van der Waerden (1933), {\it Sitzungsber. Preu{\ss}. Akad.
Wiss. Phys. Math. Kl.\/} {\bf 9}, 380--401.

\bibitem
W.L. Bade \& H. Jehle (1953), {\it Rev. Modern Phys.\/} {\bf 25}, 714.

\bibitem
R. Penrose (1960), {\it Ann. Phys.\/} {\bf 10}, 171--201.

\bibitem
A. Ashtekar (1988), {\it New Perspectives in Canonical Gravity},
Bibliopolis, Napoli.

\bibitem
M. Ferraris, M. Francaviglia \& M. Mottini (1994), {\it Rend.
Mat.\/}~(7) {\bf 14}, 457--481.

\bibitem
H.S. Ruse (1937), {\it Proc. Roy.~Soc. Edinburgh\/} A~{\bf 57}, 97--127.

\bibitem
Y. Choquet-Bruhat (1987), `Spin $1/2$ fields in arbitrary dimensions
and the Einstein-Cartan theory', in: {\it Gravitation and geometry (a
volume in honour of I.~Robinson)}, W.~Rindler \& A.~Trautman (Eds.),
Bibliopolis, Napoli, pp. 83--106.

\bibitem
M. Ferraris (1984), `Fibered Connections and Global Poincar\'e-Cartan forms
in Higher Order Calculus of Variations', in: {\it Proc. Conference on
Differential Geometry and its Applications\/} (Nov\'e M\v esto na
Morav\v e, 1983), D.~Krupka (Ed.), J.~E.~Purkyn\v e University, Brno, pp.
61--91.

\bibitem
M. Ferraris \& M. Francaviglia (1985), {\it J.~Math. Phys.\/} {\bf 24} (1),
120--124.

\bibitem
A. Komar (1959), {\it Phys. Rev.\/} {\bf 113}, 934--936.

\bibitem
D.C. Robinson (1995), {\it Class. Quant. Grav.\/} {\bf 12}, 307--315.

\bibitem
C. M{\o}ller (1961), {\it Mat. Fys. Skr. Dan. Vid. Selsk.\/} {\bf 1},
No.~10, pp. 1--50,
{\it id.} (1961), {\it Ann. Phys.} {\bf 12}, 118--133.

\bibitem
J.N. Nester (1989), {\it Phys. Lett.\/} {\bf 139A}, 112--114,
{\it id.} (1989), {\it Int. J. Mod. Phys.\/} {\bf 4A}, 1755--1772.

\bibitem
L.J. Mason \& J. Frauendiener (1990), `The Sparling 3-form, Ashtekar
Variables and Quasi-local Mass', in: {\it Twistors in Mathematics and
Physics}, London Mathematical Society Lectures Note Series {\bf 156},
T.N.~Bailey \& R.J.~Baston (Eds.),
Cambridge University Press, Cambridge, pp. 189--217.

\bibitem
L.B. Szabados (1992), {\it Class. Quant. Grav.\/} {\bf 9}, 2521--2541.

\bibitem
D. Canarutto \& A. Jadczyk (1998), `Fundamental geometric structures
for the Dirac equation in GR',
{\it Acta Applicandae Mathematicae}, {\bf 51} (1), 59-92.

\bibitem
R. Arnowitt, S. Deser \& C.W. Misner (1962), `The Dynamics of General
Relativity', in: {\it Gravitation: An Introduction to Current
Research}, L.~Witten (Ed.), Wiley, New York, pp. 227--265.

\bibitem
M. Ferraris \& M. Francaviglia (1988), `Remarks on the Energy of the
Gravitational Field', in: {\it Proc. 8th Italian Conference on General
Relativity and Gravitational Physics}, M.~Cerdonio, R.~Cianci,
M.~Francaviglia \& M.~Toller (Eds.), World Scientific, Singapore, pp.
183--196.

\bibitem
R. Penrose (1982), {\it Proc. R.~Soc. Lond.\/} A~{\bf 381}, 53--62.

\bibitem
K.P. Tod (1983), {\it Proc. R.~Soc. Lond.\/} A~{\bf 388}, 457--477.

\bibitem
A. Dougan (1992), {\it Class. Quantum Grav.}~{\bf 9}, 2461--2475.

\bibitem
P. Matteucci (1997), {\it Energia del campo gravitazionale
nell'ipotesi di simmetria sferica}, Universit\`a degli Studi di
Torino, Thesis.

}
%
%
%
\end